\documentclass[a4paper]{article}
\usepackage{graphicx}
\usepackage{amsmath}

\begin{document}

\title{Perturbation theory of von Neumann Entropy }
\author{Xiao-yu Chen \\
{\small {\ College of Information and Electronic Engineering ,Zhejiang
Gongshang University, Hangzhou, 310018, China}}}
\date{}
\maketitle

\begin{abstract}
In quantum information theory, von Neumann entropy plays an important role.
The entropies can be obtained analytically only for a few states. In
continuous variable system, even evaluating entropy numerically is not an
easy task since the dimension is infinite. We develop the perturbation
theory systematically for calculating von Neumann entropy of non-degenerate
systems as well as degenerate systems. The result turns out to be a
practical way of the expansion calculation of von Neumann entropy.
\end{abstract}

\section{Introduction}

In quantum information theory, von Neumann entropy of a state appears in
many basic theorems such as quantum source coding \cite{Schumacher}, quantum
channel coding of classical information \cite{Holevo} \cite{SchuWest},
quantum channel coding of quantum information \cite{Devetak} \cite{Barnum}.
The later two are problems of channel capacities, which are maxima of Holevo
quantity and coherent information, respectively (usually the regulation
procedure should be taken). The problems can be reduced to the calculation
of von Neumann entropy. The basic method to obtain the von Neumann entropy
of a state is to calculate its spectrum. The spectrum can seldom be obtained
for a given state. Meanwhile numeric calculation of the spectrum may
encounter problems for continuous variable system which has an infinite
dimensional Hilbert space. Hence a perturbation theory for the von Neumann
entropy of a state is needed. While quantum mechanics provides us the theory
of lower order perturbation to the energy levels, here we will develop a
systematical theory of the entropy perturbation up to any order of precision.

\section{The perturbation to the entropy of a non-degenerate system}

\subsection{The perturbation of entropy via eigenvalues}

A non-degenerate quantum system is a density matrix $\rho _0$ with its
eigenvalues $E_n\neq E_m$ for all $n\neq m$, where $n$ $($or $m)$ is the
number (or vector number) specifying the quantum eigenstate. The usual
quantum perturbation theory for the non-degenerate systems gives the first
and second order eigenvalue perturbations
\begin{eqnarray}
E_n^{(1)} &=&H_{nn}, \\
E_n^{(2)} &=&\sum_{m\neq n}\frac{\left| H_{nm}\right| ^2}{E_n-E_m},
\end{eqnarray}
when the perturbation to the density matrix is the matrix $H,$ where $%
H_{nm}=\left\langle \psi _n\right| H\left| \psi _m\right\rangle ,$ with $%
\left| \psi _n\right\rangle $ the eigenvector of the density operator $\rho
_0$ corresponding to the eigenvalue $E_n$ $.$ As usual, the perturbed
density matrix can be written as $\rho =\rho _0+\varepsilon H.$ Since $\rho $
and $\rho _0$ are density matrices, we have $H^{\dagger }=H,$and $TrH=0.$
Thus we have $\sum_nE_n^{(1)}=0,$and it is clear that $\sum_nE_n^{(2)}=0.$
The total eigenvalue up to the second order is $E_n^{(t)}=E_n+\varepsilon
E_n^{(1)}+\varepsilon ^2E_n^{(2)}.$ The entropy of the state up to the
second order perturbation is $S(\rho )=-Tr\rho \log \rho
=-\sum_nE_n^{(t)}\log E_n^{(t)}+o(\varepsilon ^2).$ Thus we have $S(\rho
)=S(\rho _0)+\varepsilon \frac{dS(\rho _0)}{d\varepsilon }+\frac
12\varepsilon ^2\frac{d^2S(\rho _0)}{d\varepsilon ^2}+o(\varepsilon ^2)$
with
\begin{eqnarray}
\frac{dS(\rho _0)}{d\varepsilon } &=&-\sum_nH_{nn}\log E_n  \label{wq1} \\
\frac{d^2S(\rho _0)}{d\varepsilon ^2} &=&-\sum_n\frac{H_{nn}^2}{E_n}%
-2\sum_nE_n^{(2)}\log E_n  \label{wq2}
\end{eqnarray}

\subsection{The expansion formula of the entropy}

For $a>0,$ we have $\log a=\int_0^\infty \frac{at-1}{a+t}\frac{dt}{1+t^2}.$
Similarly, for a positive operator $A,$ we have $\log A=\int_0^\infty \frac{%
At-1}{A+t}\frac{dt}{1+t^2}.$\cite{Vedral} Thus $\log A-\log
B=-\lim_{M\rightarrow \infty }\int_0^M(\frac 1{A+t}-\frac 1{B+t})dt$ for
positive operators $A$ and $B$. $\frac{dS(\rho _0)}{d\varepsilon }%
=-\lim_{\varepsilon \rightarrow 0}\frac 1\varepsilon Tr[\rho \log \rho -\rho
_0\log \rho _0]=-\lim_{\varepsilon \rightarrow 0}\frac 1\varepsilon Tr[\rho
_0(\log \rho -\log \rho _0)]-Tr[H\log \rho _0],$ Note that $\frac
1{A+t}=\frac 1t\sum_{k=0}^\infty (-1)^k(\frac At)^k,$ we have
\begin{equation}
\frac{dS(\rho _0)}{d\varepsilon }=\int_0^\infty Tr\{\rho _0(\rho
_0+t)^{-1}H(\rho _0+t)^{-1}\}dt-Tr[H\log \rho _0].
\end{equation}
Evaluating in the eigenbasis of $\rho _0,$ it reads
\begin{eqnarray}
\frac{dS(\rho _0)}{d\varepsilon } &=&\sum_n\int_0^\infty
E_n(E_n+t)^{-2}H_{nn}dt-\sum_nH_{nn}\log E_n  \nonumber \\
&=&-\sum_nH_{nn}\log E_n,  \label{wq3}
\end{eqnarray}
where $TrH=0$ has been used. It coincides with Eq.(\ref{wq1}), the result
derived from the perturbation of the eigenvalues.

The second derivative of the entropy at $\rho _0$ is $\frac{d^2S(\rho _0)}{%
d\varepsilon ^2}=-\lim_{\varepsilon \rightarrow 0}\frac 1{\varepsilon
^2}Tr[\rho _{+}\log \rho _{+}+\rho _{-}\log \rho _{-}-2\rho _0\log \rho _0],$
where $\rho _{\pm }=\rho _0\pm \varepsilon H.$ The derivative can be
rewritten as $P_1+P_2,$ the two parts are $P_1=-\lim_{\varepsilon
\rightarrow 0}\frac 1{\varepsilon ^2}Tr\rho _0(\log \rho _{+}+\log \rho
_{-}-2\log \rho _0)=$ $2\int_0^\infty Tr\rho _0(\rho _0+t)^{-1}H$ $(\rho
_0+t)^{-1}H(\rho _0+t)^{-1}dt$ and $P_2=-\lim_{\varepsilon \rightarrow
0}\frac 1\varepsilon TrH(\log \rho _{+}-\log \rho _{-})=$ $-2\int_0^\infty
TrH(\rho _0+t)^{-1}H(\rho _0+t)^{-1}dt.$ With the eigenbasis of $\rho _0,$
we have
\begin{eqnarray}
\frac{d^2S(\rho _0)}{d\varepsilon ^2} &=&-2\sum_{n,m}\int_0^\infty \frac{%
t\left| H_{nm}\right| ^2}{(E_n+t)^2(E_m+t)}dt  \nonumber \\
&=&-\sum_n\frac{\left| H_{nn}\right| ^2}{E_n}-2\sum_{n,m\neq n}\frac{\left|
H_{nm}\right| ^2E_m}{(E_n-E_m)^2}\log \frac{E_m}{E_n}  \nonumber \\
&=&-\sum_n\frac{\left| H_{nn}\right| ^2}{E_n}-2\sum_{n,m\neq n}\frac{\left|
H_{nm}\right| ^2}{E_n-E_m}\log E_n  \label{wq4}
\end{eqnarray}
Note that Eq.(\ref{wq2}) and Eq. (\ref{wq4}) are strictly the same, and the
non-degenerate condition is used.

The $n^{th}$ derivative of the entropy can also be carried out. From the
definition of the derivative, we have $\frac{d^nS(\rho _0)}{d\varepsilon ^n}%
=-\lim_{\varepsilon \rightarrow 0}\frac 1{\varepsilon ^n}\sum_{l=0}^nS(\rho
_0+l\varepsilon )\binom ln(-1)^l=P_3+P_4$, with $P_3=-\lim_{\varepsilon
\rightarrow 0}\frac 1{\varepsilon ^n}\sum_{l=0}^nTr\rho _0\log (\rho
_0+l\varepsilon )\binom ln(-1)^l=n!$ $\int_0^\infty Tr\rho _0(\rho
_0+t)^{-1}[H(\rho _0+t)^{-1}]^n$ $dt$ for $n>0.$ $P_4=-\lim_{\varepsilon
\rightarrow 0}\frac 1{\varepsilon ^n}\sum_{l=0}^nTr[\varepsilon lH\log (\rho
_0+l\varepsilon )]\binom ln(-1)^l=$ $-n!\int_0^\infty Tr[H(\rho
_0+t)^{-1}]^n $ $dt$ for $n>1$ ( see Appendix A). Thus we have
\begin{equation}
\frac{d^nS(\rho _0)}{d\varepsilon ^n}=-(-1)^nn!\int_0^\infty Tr\{t(\rho
_0+t)^{-1}[H(\rho _0+t)^{-1}]^n\}dt.
\end{equation}
The Taylor expansion formula for the entropy of a state $\rho =\rho
_0+\varepsilon H$ will be
\begin{eqnarray}
S(\rho ) &=&\sum_{n=0}^\infty \frac{\varepsilon ^n}{n!}\frac{d^nS(\rho _0)}{%
d\varepsilon ^n}  \nonumber \\
&=&S(\rho _0)-\varepsilon Tr[H\log \rho _0]-\sum_{n=2}^\infty (-\varepsilon
)^n\int_0^\infty Tr\{t(\rho _0+t)^{-1}[H(\rho _0+t)^{-1}]^n\}dt.  \label{wq5}
\end{eqnarray}

We calculate the third and the fourth order perturbation for the case of $%
H_{nn}=0,$ the entropy up to fourth order perturbation is

\begin{equation}
S(\rho )=S(\rho _0)-\varepsilon ^2\sum_{n,m\neq n}\frac{\left| H_{nm}\right|
^2}{E_n-E_m}\log E_n+\varepsilon ^3Q_3-\varepsilon
^4(Q_{41}+Q_{42}+Q_{43})+o(\varepsilon ^4)
\end{equation}
where $Q_3,Q_{41},Q_{42}$ and $Q_{43}$ are given in Appendix B.

\subsection{Perturbation to the one-mode thermal state}

Assuming that a one-mode thermal state is perturbed, the perturbed
characteristic function is
\begin{equation}
\chi (\mu )=\chi _T(\mu )[1+\varepsilon (\alpha ^{*}\mu -\alpha \mu ^{*})],
\end{equation}
where $\chi _T(\mu )=\exp [-(N+\frac 12)\left| \mu \right| ^2]$ is the
characteristic function of the one-mode thermal state, with $N$ the average
photon number of the state. The density operator can be obtained with $\rho
=\int \frac{d^2\mu }\pi \chi (\mu )D(-\mu ),$ where $D(\mu )=\exp [\mu
a^{\dagger }-\mu ^{*}a]$ is the displacement operator, $a^{\dagger }$and $a$
are the creation and annihilation operator of the system, respectively. By
the method of integral within ordered product operator, the perturbed
density operator is
\begin{equation}
\rho =\rho _T+\varepsilon (1-v)(\alpha a^{\dagger }\rho _T+\alpha ^{*}\rho
_Ta).
\end{equation}
where $\rho _T=(1-v)\sum_{n=0}^\infty v^n\left| n\right\rangle \left\langle
n\right| $ is the unperturbed one-mode thermal state with $v=N/(N+1).$ It is
clear that in the eigenbasis of $\rho _T$ the diagonal elements of the
perturbation $H=(1-v)(\alpha a^{\dagger }\rho _T+\alpha ^{*}\rho _Ta)$ is
null, thus in the evaluation of Eq.(\ref{wq5}) we only need to consider the
off-diagonal elements of $H.$ Evaluating Eq. (\ref{wq5}) in the basis of $%
\rho _T,$up to $\varepsilon ^4$, for a null-diagonal $H$, the entropy will
be
\begin{equation}
S(\rho )=S(\rho _T)-\varepsilon ^2\sum_{n,m\neq n}\frac{\left| H_{nm}\right|
^2}{E_n-E_m}\log E_n+\varepsilon ^3Q_3-\varepsilon
^4(Q_{41}+Q_{42}+Q_{43})+o(\varepsilon ^4)  \label{wq6}
\end{equation}
In our case, $H_{nm}=(1-v)(\alpha \sqrt{n}E_{n-1}\delta _{n,m+1}+\alpha
^{*}E_n\sqrt{n+1}\delta _{n,m-1}).$ Thus $Q_3=Q_{41}=0$ due to the structure
of $H,$ and $Q_{43}=\left| \alpha \right| ^4[\frac{(1+v)^2}{2v}+\frac{1+v}{%
1-v}\log v],$ $Q_{42}=-2\left| \alpha \right| ^4[1+\frac{1+v^2}{1-v^2}\log
v].$ The entropy of the perturbed state is
\begin{equation}
S(\rho )=S(\rho _T)-2\varepsilon ^2\left| \alpha \right| ^2\log \frac
1v-\varepsilon ^4\left| \alpha \right| ^4[\frac{(1-v)^2}{2v}+\frac{1-v}{1+v}%
\log \frac 1v]+o(\varepsilon ^4).  \label{wq10}
\end{equation}

\section{The perturbation to the entropy of a degenerate system}

\subsection{The entropy perturbation up to second order}

Note that Eq.(\ref{wq5}) is an overall result regardless of the eigenvalue
structure of the state $\rho _0.$ The difference between the degenerate
system and non-degenerate system comes when we evaluate the entropy. We now
classify the eigenvectors according to the eigenvalues of $\rho _0.$ Suppose
the eigenvalue $E_n$ correspond to the eigenvector set $\left\{ \left|
n,n_i\right\rangle ,\left| i=1,\ldots n_d\right. \right\} $, these
eigenvectors span the subspace of dimension $n_d$ for $E_n$. Define $%
n_d\times m_d$ matrix $H_{\mathbf{nm}}$ with its entries being
\begin{equation}
H_{nn_i,mm_j}=\left\langle n,n_i\right| H\left| m,m_j\right\rangle .
\end{equation}
Consider the first order derivative to the entropy, we have $\alpha \alpha $%
\begin{eqnarray}
\frac{dS(\rho _0)}{d\varepsilon } &=&\sum_n\int_0^\infty E_n(E_n+t)^{-2}TrH_{%
\mathbf{nn}}dt-\sum_nTrH_{\mathbf{nn}}\log E_n  \nonumber \\
&=&-\sum_nTrH_{\mathbf{nn}}\log E_n.  \label{wq7}
\end{eqnarray}
Since $TrH_{\mathbf{nn}}=\sum_{i=1}^{n_d}\left\langle n,n_i\right| H\left|
n,n_i\right\rangle $ is an invariant in the subspace, the first order
perturbation to the entropy can be written as $-\sum_kH_{kk}\log E_k,$ where
$k$ is the unified label for all the distinct eigenvectors (for some $%
k^{\prime }\neq k,$ we may have $E_{k^{\prime }}=E_k$). Thus the degenerate
of the eigenvalues will not affect the expression of the first order
perturbation of the entropy. With the notation of $H_{\mathbf{nm}},$ it
follows the second order derivative to the entropy
\begin{equation}
\frac{d^2S(\rho _0)}{d\varepsilon ^2}=-\sum_n\frac{Tr[H_{\mathbf{nn}}H_{%
\mathbf{nn}}^T]}{E_n}-2\sum_{n,m\neq n}\frac{Tr[H_{\mathbf{nm}}H_{\mathbf{nm}%
}^T]}{E_n-E_m}\log E_n  \label{wq8}
\end{equation}
where $Tr[H_{\mathbf{nn}}H_{\mathbf{nn}}^T]=Tr[H_{\mathbf{nn}}H_{\mathbf{nn}%
}^{*}]=\sum_{i,j}\left| H_{nn_i,nn_j}\right| ^2$ is the summation of the
absolute square of all entries of the matrix $H_{\mathbf{nn}},$not the
summation of the absolute square of diagonal entries of the matrix $H_{%
\mathbf{nn}}$.

\subsection{Perturbation to the two-mode thermal state}

Let $\rho _T=(1-v)\sum_{n=0}^\infty v^n\left| n\right\rangle \left\langle
n\right| $ be single-mode thermal state, the direct product of $\rho _T$
with itself will result a two-mode $\rho _2=$ $\rho _T\times \rho _T.$ The
characteristic function of $\rho _2$ is $\chi _2(\mu _1,\mu _2)=\exp
[-(N+\frac 12)(\left| \mu _1\right| ^2+\left| \mu _2\right| ^2)].$ We
consider the perturbed characteristic function $\chi (\mu _1,\mu _2)=\chi
_2(\mu _1,\mu _2)[1+\varepsilon (\alpha \mu _1\mu _2+\alpha ^{*}\mu
_1^{*}\mu _2^{*})].$ The perturbed density operator is $\rho =\rho
_2+\varepsilon H$, with
\begin{equation}
H=(1-v)^2(\alpha ^{*}a_1^{\dagger }a_2^{\dagger }\rho _2+\alpha \rho
_2a_1a_2),
\end{equation}
where $a_i^{\dagger }$ and $a_i$ are the creation and annihilation operator
for the two modes, respectively. The state $\rho _2=(1-v)^2\sum_{n=0}^\infty
\sum_{j=0}^nv^n\left| j,n-j\right\rangle \left\langle j,n-j\right| .$ For
any given $n>0,$ the state is $(n+1)$-fold degenerate. In the eigenbasis of $%
\rho _2$, the entries of matrix $H_{\mathbf{nm}}$ are
\begin{eqnarray}
\left\langle j,n-j\right| H\left| k,m-k\right\rangle &=&(1-v)^2[\alpha
^{*}\delta _{j,k+1}\delta _{n,m+2}E_{n-2}\sqrt{j(n-j)}  \nonumber \\
&&+\alpha \delta _{j,k-1}\delta _{n,m-2}E_n\sqrt{(j+1)(n-j+1)}],
\end{eqnarray}
where $E_n=(1-v)^2v^n$ is the eigenvalue of $\rho _2.$ Clearly, $H_{\mathbf{%
nn}}$ now is a zero matrix, so that $TrH_{\mathbf{nn}}=Tr[H_{\mathbf{nn}}H_{%
\mathbf{nn}}^T]=0$. We only need to evaluate the second term of Eq. (\ref
{wq8}). We have
\begin{eqnarray}
\sum_{n,m\neq n}\frac{Tr[H_{\mathbf{nm}}H_{\mathbf{nm}}^T]}{E_n-E_m}\log E_n
&=&(1-v)^4\left| \alpha \right| ^2\sum_n[\frac{E_{n-2}^2}{E_n-E_{n-2}}%
\sum_{j=0}^nj(n-j)  \nonumber \\
&&+\frac{E_n^2}{E_n-E_{n+2}}\sum_{j=0}^n(j+1)(n-j+1)]\log v^n  \nonumber \\
&=&-\frac{2\left| \alpha \right| ^2}{1+v}\log \frac 1v.
\end{eqnarray}
Thus the entropy of the perturbed state will be
\begin{equation}
S(\rho )=2S(\rho _T)-\frac{\varepsilon ^2\left| \alpha \right| ^2}{1+v}\log
\frac 1v+o(\varepsilon ^2).
\end{equation}

\section{Density operator expansion}

In the calculation of the coherent information, even if we assume that the
input state undergo a simple form perturbation $\rho =\rho _0+\varepsilon H$%
, the output state $\mathcal{E}(\rho )$ and the combined output state $(%
\mathcal{I}\otimes \mathcal{E}$ $)\left| \Psi \right\rangle \left\langle
\Psi \right| $ may have a more complicated perturbation structures by the
application of channel map $\mathcal{E}$ and the purification from $\rho $
to $\left| \Psi \right\rangle $. So that we need to consider a more generic
form of perturbation:
\begin{equation}
\rho =\rho _0+\varepsilon H^{(1)}+\varepsilon ^2H^{(2)}+\varepsilon
^3H^{(3)}+\cdots .
\end{equation}
It is reasonable to require that $TrH^{(n)}=0$ is true for each $n$. The
entropy perturbation can be obtained in due course, which is
\begin{eqnarray}
S(\rho ) &=&S(\rho _0)-\sum_{n=1}^\infty \varepsilon ^nTr[H^{(n)}\log \rho
_0]-\sum_{n=2}^\infty \varepsilon ^n\sum_{j=2}^n(-1)^j\sum_{i_1i_2\cdots
i_j}\int_0^\infty t(\rho _0+t)^{-1}H^{(i_1)}(\rho _0+t)^{-1}  \nonumber \\
&&\times H^{(i_2)}(\rho _0+t)^{-1}\cdots H^{(i_j)}(\rho _0+t)^{-1}\delta
_{i_1+i_2+\cdots +i_j,n}.
\end{eqnarray}
We may express the second and the third order derivatives of the entropy in
more explicit forms:
\begin{eqnarray}
\frac 1{2!}\frac{d^2S(\rho _0)}{d\varepsilon ^2} &=&-Tr\int_0^\infty t(\rho
_0+t)^{-1}H^{(1)}(\rho _0+t)^{-1}H^{(1)}(\rho _0+t)^{-1}dt  \nonumber \\
&&-Tr[H^{(2)}\log \rho _0],  \label{wq9}
\end{eqnarray}
\begin{eqnarray}
\frac 1{3!}\frac{d^3S(\rho _0)}{d\varepsilon ^3} &=&Tr\int_0^\infty t(\rho
_0+t)^{-1}[H^{(1)}(\rho _0+t)^{-1}]^3dt  \nonumber \\
&&-2Tr\int_0^\infty t(\rho _0+t)^{-1}H^{(1)}(\rho _0+t)^{-1}H^{(2)}(\rho
_0+t)^{-1}dt  \nonumber \\
&&-Tr[H^{(3)}\log \rho _0].
\end{eqnarray}

To verify the entropy expansion formula, we consider the displacement of
thermal state, $\rho _d=D(\varepsilon \alpha )\rho _TD^{\dagger
}(\varepsilon \alpha ).$ The characteristic function of the state is $\chi
_d\left( \mu \right) =Tr[\rho _dD(\mu )]=$ $Tr[\rho _TD^{\dagger
}(\varepsilon \alpha )D(\mu )D(\varepsilon \alpha )]=\chi _T(\mu )\exp
[\varepsilon (\alpha ^{*}\mu -\alpha \mu ^{*})].$ Since the displacement
operator is unitary, the entropy will not be changed by the displacement
operation. Thus $S(\rho _d)=S(\rho _T)$, the entropy $S(\rho _d)$ is not a
function of $\varepsilon .$ If we expand the entropy of $\rho _d$ with
respect to $\varepsilon ,$ then the perturbation up to any order should be
zero. Expanding with respect to $\varepsilon ,$ the characteristic function
is
\begin{equation}
\chi _d\left( \mu \right) =\chi _T(\mu )[1+\varepsilon (\alpha ^{*}\mu
-\alpha \mu ^{*})+\frac 1{2!}\varepsilon ^2(\alpha ^{*}\mu -\alpha \mu
^{*})^2+\cdots ].
\end{equation}
The state is $\rho _d=\rho _T+$ $\varepsilon H^{(1)}+\varepsilon
^2H^{(2)}+\cdots ,$ with
\begin{equation}
H^{(2)}=\frac 1{2!}(1-v)^2[\alpha ^2a^{\dagger 2}\rho _T+\rho _T\alpha
^{*2}a^2+2\left| \alpha \right| ^2a^{\dagger }\rho _Ta]-(1-v)\left| \alpha
\right| ^2\rho _T.  \label{wq11}
\end{equation}
The $H^{(1)}$ term has been treated in the former section, which give rise
to the first term of the right hand side of Eq. (\ref{wq9}) the value $%
2\left| \alpha \right| ^2\log v$ . And by direct calculation we have
\begin{equation}
-Tr[H^{(2)}\log \rho _0]=-2\left| \alpha \right| ^2\log v.
\end{equation}
So $\frac{d^2S(\rho _0)}{d\varepsilon ^2}=0.$ The first and the third
derivatives of the entropy should be $0$ considering the symmetry of the
system.

\section{Discussion and Conclusion}

In the examples presented, we obtain the density operator perturbation via
the expansion of the characteristic function for continuous variable
systems. This has some merits. First of all, since the characteristic
function is a c-number function, its expansion is simply Taylor expansion.
We do not consider the Laurent expansion of the complex function since for
any state we have $\chi (0)$ $=1.$ The characteristic function is defined as
$\chi (\mu ) $ $=Tr[\rho D(\mu )],$ thus $\chi (0)$ $=Tr[\rho D(0)]=Tr[\rho
]=1.$ Secondly, we expand the characteristic function as $\chi (\mu )=\chi
_0(\mu )[1+\sum_{n=1}^\infty \varepsilon ^n\sum_{i=0}^nc_{i,n-i}\mu ^i\mu
^{*j}]$ for the single-mode system (the multi-mode expansion can be obtained
in due course). The unperturbed characteristic function $\chi _0(\mu )$ is
reserved as an integral nuclear for each order of perturbations. Thus the $%
n- $th perturbation $H^{(n)}=\int \frac{d^2\mu }\pi \chi _0(\mu
)\sum_{i=0}^nc_{i,n-i}\mu ^i\mu ^{*j}D(-\mu )$ can be obtained with the
difficulty of non-integrable. This form of expansion also implies that $%
Tr[H^{(n)}]=0,$ which is not apparent at the first sight when we write down,
for example, $H^{(2)}$ in Eq. (\ref{wq11}), although it can be proved by
simple calculation.

One of the obstacle that may be encountered in the entropy perturbation
calculation is the $0$ eigenvalues of the unperturbed system. For instance,
we consider the perturbation to a pure state, the unperturbed eigenvalues
are $0$ and $1.$ This can be partially overcome by decomposing the
perturbation matrix $H$ as diagonal and off-diagonal part according to the
eigensystem of the unperturbed density matrix $\rho _0.$ Suppose $\rho =\rho
_0+\varepsilon (H_0+H^{\prime }),$ with $H_0$ being diagonal and $H^{\prime
} $ being off-diagonal in the eigenbasis of $\rho _0.$ So $\rho _0$ and $H_0$
can be diagonalized simultaneously. Thus the perturbed state can be written
as $\rho =\rho _0^{\prime }+\varepsilon H^{\prime }$ with a new expansion
base state $\rho _0^{\prime }=$ $\rho _0+\varepsilon H_0.$ Usually the
eigenvalues of $\rho _0^{\prime }$ will differ from $0$.

We have derived the entropy perturbation formula for degenerate and
non-degenerate system. Up to any give order of perturbation of the density
operator, the entropy perturbation was written in a concise form via
operator integral, the integral then can be evaluated in the eigenbasis of
the unperturbed density operator of the system.

\section*{Appendix A: The n-th order derivative of entropy}

$\frac{d^nS(\rho _0)}{d\varepsilon ^n}==P_3+P_4$, with

\begin{eqnarray*}
P_3 &=&-\lim_{\varepsilon \rightarrow 0}\frac 1{\varepsilon
^n}\sum_{l=0}^nTr\rho _0\log (\rho _0t+l\varepsilon )\binom ln(-1)^{n-l} \\
&=&-\lim_{\varepsilon \rightarrow 0}\frac{(-1)^n}{\varepsilon ^n}%
\sum_{l=0}^n(-1)^l\binom lnTr\rho _0\int_0^\infty \frac{-dt}{\rho
_0+l\varepsilon +t} \\
&=&\lim_{\varepsilon \rightarrow 0}\frac{(-1)^n}{\varepsilon ^n}%
\sum_{l=0}^n(-1)^l\binom lnTr\rho _0\int_0^\infty \frac{dt}%
t\sum_{m=0}^\infty (-\frac{\rho _0+l\varepsilon H}t)^m \\
&=&\lim_{\varepsilon \rightarrow 0}\frac{(-1)^n}{\varepsilon ^n}%
\sum_{l=0}^n(-1)^l\binom lnTr\rho _0\int_0^\infty \frac{dt}%
t\sum_{m=n}^\infty \sum_{i_1,\cdots i_n}(-\frac{\rho _0}t)^{i_1}\frac{%
l\varepsilon H}t\cdots (-\frac{\rho _0}t)^{i_n}\frac{l\varepsilon H}t(-\frac{%
\rho _0}t)^{m-n-\sum_{j=0}^ni_j}(-1)^n \\
&=&\sum_{l=0}^n(-1)^l\binom lnl^n\int_0^\infty Tr\rho _0(\rho
_0+t)^{-1}[H(\rho _0+t)^{-1}]^ndt \\
&=&(-1)^nn!\int_0^\infty Tr\rho _0(\rho _0+t)^{-1}[H(\rho _0+t)^{-1}]^ndt.
\end{eqnarray*}
The last equation cames from the identity $\sum_{l=0}^n(-1)^l\binom
lnl^n=(-1)^nn!.$ In the fourth equation we have used the fact that $%
\sum_{l=0}^n(-1)^l\binom lnl^m=0$ for all $m<n.$ We now prove this two
results. Assuming $g(x)=(1-x)^n=\sum_{l=0}^n(-1)^l\binom lnx^l,$ let $%
f_m(x)=\left[ x\frac d{dx}\right] ^mg(x)=\sum_{l=0}^n(-1)^l\binom lnl^mx^l,$
then $f_m(x)=\sum_{k=1}^mc_k^{(m)}x^k\frac{d^kg(x)}{dx^k},$ where $c_k^{(m)}$
is the coefficient which obey the recursive relation $%
c_k^{(m+1)}=kc_k^{(m)}+c_{k-1}^{(m)}.$ We always have $c_m^{(m)}=1.$ Note
that $\frac{d^kg(0)}{dx^k}=0,$ for all $k<n$ , and $\frac{d^kg(0)}{dx^k}%
=(-1)^nn!,$ for $k=n.$ Thus we have $\sum_{l=0}^n(-1)^l\binom
lnl^n=f_n(1)=(-1)^nn!$ and $\sum_{l=0}^n(-1)^l\binom lnl^m=f_m(1)=0$ for all
$m<n$.

$P_4$ can be obtained in the very similar way.

\section*{Appendix B: The third and fourth order off-diagonal perturbation}

When the perturbation is off-diagonal in the eigenbasis of the unperturbed
state, we have $H_{nn}=0.$ The contribution to the third order perturbation
should be
\[
Q_3=\int_0^\infty Tr\{t(\rho _0+t)^{-1}[H(\rho _0+t)^{-1}]^3\}dt.
\]
Evaluating in the basis of $\rho _0,$ we have
\[
Q_3=\sum_{n,m,k}\int_0^\infty \frac{tH_{nm}H_{mk}H_{kn}}{%
(E_n+t)^2(E_m+t)(E_k+t)}dt.
\]
The summation should be taken for the case that $n,m$ and $k$ are all
different since $H_{nn}=0.$ After integrating and considering the
permutation symmetry, we have

\[
Q_3=\sum_{n<m<k}\frac{2Re(H_{nm}H_{mk}H_{kn})}{E_m-E_k}[\frac{\log (E_n/E_k)%
}{(E_n-E_k)}-\frac{\log (E_n/E_m)}{E_n-E_m}],
\]
The contribution to the fourth order perturbation is
\[
Q_4=\sum_{n,m,l,k}\int_0^\infty \frac{tH_{nm}H_{ml}H_{lk}H_{kn}}{%
(E_n+t)^2(E_m+t)(E_l+t)(E_k+t)}dt.
\]
When all $n,m,l,k$ are different with each other, the integral will be

\begin{eqnarray*}
&&Q_{41}=%
\sum_{n<m<l<k}2Re(H_{nm}H_{ml}H_{lk}H_{kn}+H_{nm}H_{ml}H_{lk}H_{kn}+H_{nm}H_{ml}H_{lk}H_{kn})
\\
&&[\frac{\log (E_m/E_n)}{(E_m-E_n)(E_m-E_l)(E_m-E_k)}+\frac{\log (E_l/E_n)}{%
(E_l-E_n)(E_l-E_m)(E_l-E_k)} \\
&&+\frac{\log (E_k/E_n)}{(E_k-E_n)(E_k-E_m)(E_k-E_l)}],
\end{eqnarray*}
where permutation symmetry are considered. When two of $n,m,l,k$ are equal,
that is (i) $n=l\neq m\neq k$ and (ii) $m=k\neq n\neq l,$ the integral will
be
\[
Q_{42}=\sum_{n,m,k}\int_0^\infty \frac{t\left| H_{nm}H_{nk}\right| ^2}{%
(E_n+t)^3(E_m+t)(E_k+t)}dt+\sum_{n,l,m}\int_0^\infty \frac{t\left|
H_{mn}H_{ml}\right| ^2}{(E_n+t)^2(E_m+t)^2(E_l+t)}dt.
\]
which is
\begin{eqnarray*}
Q_{42} &=&\sum_{n<m<k}(\left| H_{nm}H_{nk}\right| ^2+\left|
H_{mn}H_{mk}\right| ^2+\left| H_{kn}H_{km}\right| ^2) \\
&&[\frac 1{E_n(E_n-E_m)(E_n-E_k)}+\frac{\log (E_m/E_n)}{(E_m-E_n)^2(E_m-E_k)}%
+\frac{\log (E_k/E_n)}{(E_k-E_n)^2(E_k-E_m)}].
\end{eqnarray*}
When $n=l$ and $m=k,$ we have the integral
\[
Q_{43}=\sum_{n,m}\int_0^\infty \frac{t\left| H_{nm}\right| ^4}{%
(E_n+t)^3(E_m+t)^2}dt,
\]
which is
\[
Q_{43}=\sum_{n<m}\left| H_{nm}\right| ^4[\frac 1{2(E_n-E_m)^2}(\frac
1{E_n}+\frac 1{E_m})-\frac 1{(E_n-E_m)^3}\log \frac{E_n}{E_m}].
\]

\end{document}